\documentclass{emulateapj}
\usepackage{epsfig}
\usepackage{longtable}
\def\gsim{\lower 2pt \hbox{$\, \buildrel {\scriptstyle >}\over
{\scriptstyle \sim}\,$}}
\def\lsim{\lower 2pt \hbox{$\, \buildrel {\scriptstyle <}\over
{\scriptstyle \sim}\,$}}

\def\fuse{{\em FUSE}}

\def\chandra{{\em Chandra}}
\def\ovii{O{\small~VII}}
\def\oviii{O{\small~VIII}}
\def\ovi{O{\small~VI}}
\def\oi{O{\small~I}}
\def\oii{O{\small~II}}
\def\oiii{O{\small~III}}
\def\neix{Ne{\small~IX}}

\newcommand{\etal}{et al.~}

\begin{document}
\slugcomment{\em Accepted for publication in the  Astrophysical Journal}
\title{Warm-Hot Gas in and around the Milky Way: \\ Detection and
Implications of \ion{O}{7} Absorption toward LMC X-3}
\author{Q. D. Wang\altaffilmark{1}, Y. Yao\altaffilmark{1},
T. M. Tripp\altaffilmark{1}, T.-T. Fang\altaffilmark{2},
W. Cui\altaffilmark{3}, \\ F. Nicastro\altaffilmark{4,5}, 
S. Mathur\altaffilmark{6}, R. J. Williams\altaffilmark{6}, 
L. Song\altaffilmark{1}, R. Croft\altaffilmark{7}}

\altaffiltext{1}{Astronomy Department, University of Massachusetts, Amherst, MA 01003,USA, wqd@astro.umass.edu, yaoys@astro.umass.edu}
\altaffiltext{2}{Astronomy Department, University of California, Berkeley, USA
}
\altaffiltext{3}{Physics  Department, Purdue University, USA}
\altaffiltext{4}{Harvard-Smithsonian Center for Astrophysics, USA}
\altaffiltext{5}{Instituto de Astronomia, Universidad Nacional 
  Autonoma de Mexico, Mexico, D.F., Mexico}
\altaffiltext{6}{Department of Astronomy, The Ohio State University, Columbus, USA}
\altaffiltext{7}{Physics  Department, Carnegie Mellon University, USA}

\shortauthors{Wang et al.}
\shorttitle{Galactic \ovii-bearing Gas}

\begin{abstract}
X-ray absorption lines of highly-ionized species such as \ion{O}{7} at
about zero redshift have been firmly detected in the spectra of several
active galactic nuclei. However, the location of the absorbing
gas remains a subject of debate. To separate the Galactic and
extragalactic contributions to the absorption, we have obtained {\sl
Chandra} LETG-HRC and {\sl Far Ultraviolet Spectroscopic Explorer}
observations of the black hole X-ray binary LMC X--3. We clearly
detect the \ion{O}{7} K$\alpha$ absorption line with an equivalent
width of 20(14, 26) m\AA~(90\% confidence range).  The Ne IX K$\alpha$
absorption line is also detected, albeit marginally. A joint analysis
of these lines, together with the non-detection of the \ion{O}{7}
K$\beta$ and \ion{O}{8} K$\alpha$ lines, gives the temperature,
velocity dispersion, and hot oxygen column density as $1.3(0.7,
1.8)\times10^6$ K, $79(62, 132) {\rm~km~s^{-1}}$, and 1.9(1.2,
3.2)$\times10^{16} {\rm~cm^{-2}}$, assuming a collisional ionization
equilibrium of the X-ray-absorbing gas and a Galactic interstellar 
Ne/O number ratio of 0.18. The X-ray data allow us to place a 
95\% confidence lower limit to the Ne/O ratio as 0.14, but the upper
limit is not meaningfully constrained.  The \ion{O}{7} line centroid
and its relative shift from the Galactic \ion{O}{1} K$\alpha$
absorption line, detected in the same observations, are {\sl
inconsistent} with the systemic velocity of LMC X--3 ($+310
{\rm~km~s^{-1}}$). The far-UV spectrum shows \ion{O}{6} absorption at
Galactic velocities, but no \ion{O}{6} absorption is detected at the LMC
velocity at $> 3\sigma$ significance. The measured Galactic \ion{O}{6} column
density is higher than the value predicted from the \ion{O}{7}-bearing
gas, indicating multi-phase absorption. Both
the nonthermal broadening and the decreasing scale height with
the increasing ionization state 
further suggest an origin of the highly-ionized gas
in a supernova-driven galactic fountain. 
In addition, we estimate the warm and hot electron column densities from
our detected \oii\ K$\alpha$ line in the LMC X--3 X-ray spectra 
and from the dispersion measure of a pulsar in the LMC vicinity. 
We then infer the O/H ratio of the gas to be $\gtrsim 8 \times 10^{-5}$,
consistent with the chemically-enriched galactic fountain scenario. 
We conclude that the Galactic hot interstellar medium should in general 
substantially contribute to zero-redshift X-ray absorption lines in 
extragalactic sources.

\end{abstract}


\section{Introduction}
 Recently, several groups have reported the detection of H- and
He-like oxygen absorption lines from the local ($c z \sim 0$) hot gas
in the X-ray spectra of active galactic nuclei (AGNs) with {\sl Chandra} and
{\sl XMM-Newton} X-ray Observatories 
\citep{fang02, fang03, nic02, ras03, mck04, williams05}. The detected
column densities of highly-ionized oxygen are around $10^{16}\ \rm
cm^{-2}$, which has been suggested to be an indication for the
presence of large amounts of intergalactic ``warm-hot''
gas at temperatures around $\sim
10^6$ K. This suggestion, if confirmed, would have strong implications
for our understanding of the so-called missing baryon problem
\citep{fuk98, cen99, dave01, fang02, nic05}. In turn, by comparing nearby X-ray
absorbers to highly UV ionized absorption systems observed at various
redshifts from $z \approx$ 0.2 out to $z \gtrsim$ 4 \citep{tripp00,
reime01, simcoe02, kt97}, these data could shed light on the thermal and
chemical evolution of the Universe.
However, one critical question remains open: where are these
highly-ionized $c z \sim 0$ X-ray absorbers --- in the interstellar 
medium (ISM) of
the Galactic disk, in an extended Galactic halo, in the Local Group,
or even beyond? 

 We here report {\sl Chandra} grating observations of LMC X--3 in its
high state to discriminate between the Galactic and extragalactic
scenarios of the \ion{O}{7} absorption. Because this bright X-ray source
is in our neighboring galaxy, the Large Magellanic Cloud (LMC) at the
distance of $\sim 50$ kpc, the detection of the absorption offers a
direct measurement of the hot gas content along a line of sight
through essentially the entire Galaxy, including its halo. The 
consistency between this measurement and those in the AGN sight-lines
provides a direct constraint on the residual absorption on scales
greater than the LMC distance. We
supplement the {\sl Chandra} data with a nearly simultaneous 
far-UV observation
obtained with the {\sl Far Ultraviolet Spectroscopic Explorer (FUSE)}.

\begin{figure}[thb!]
  \centerline{
      \epsfig{figure=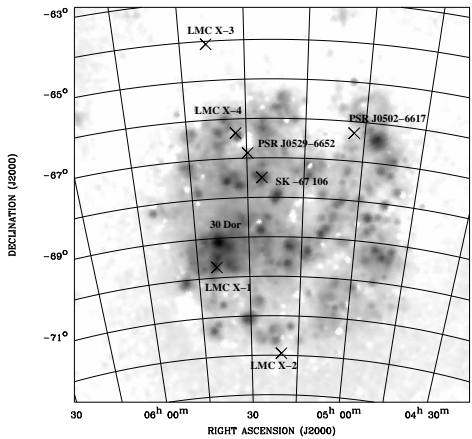,width=0.45\textwidth}
    }
  \caption{H$\alpha$ image of the LMC \citep{Fink03} with the marked 
    locations of several objects mentioned in the text.
    \label{fig:lmc_ha}}
\end{figure}

LMC X--3 is perhaps the best-suited source for this experiment.  
Among the several
bright X-ray binaries in both the Small and Large Magellanic Clouds,
LMC X--3 normally exhibits a very soft X-ray spectrum, which is
dominated by the accretion disk emission, and an exceptionally high
luminosity during its high state. Unlike most black
hole systems, LMC X--3 spends most of its time in the high
state. LMC X--3 is outside the main body of the LMC and
is about 5$^\circ$ away from the active star forming 30 Doradus
region, where LMC X--1 and two neutron star X-ray binaries X--2 and X--4 
are located (Fig.~\ref{fig:lmc_ha}).  
Therefore, the potential
absorption contribution from hot gas local to the LMC is
minimal. Finally, LMC X--3 has a systemic velocity of
$+310\pm7 {\rm~km~s^{-1}}$, comparable to the expected velocity ($+343
{\rm~km~s^{-1}}$) for the projected location in the LMC (Cowley et
al. 1983). This velocity makes it relatively straight forward to 
distinguish a spectral line that is local to the X-ray binary or to the LMC
from that associated with the Galactic disk.

We have analyzed and interpreted the \chandra\ and \fuse\ data
almost independently, but the results and conclusions found are in
nearly complete agreement. Therefore, we try to keep this independence
in the presentation of this paper.
We adopt the Galactic ISM metal abundances
given by \citet{wilms00} (e.g., the oxygen and neon number fractions are $4.90
\times 10^{-4}$ and $8.71
\times 10^{-5}$, compared to the typically assumed solar values 
$6.76\times 10^{-4}$ and $1.20 \times 10^{-4}$; \citet{grev98}); 
both the constraint on the abundances and the effect of the relative 
elemental fraction
change on our results are discussed in
\S 4. Furthermore, our spectroscopic analysis assumes that the
X-ray-absorbing gas is in a collisional ionization 
equilibrium (CIE). 
The parameter uncertainties in this paper are quoted at the 68\% and 90\% 
confidence levels for the far-UV and X-ray measurements, respectively, unless otherwise noted.

\section{Observations and Data Reduction}

\subsection{{\it Chandra} Observations}

\begin{figure}[htb!]
  \centerline{
    \epsfig{figure=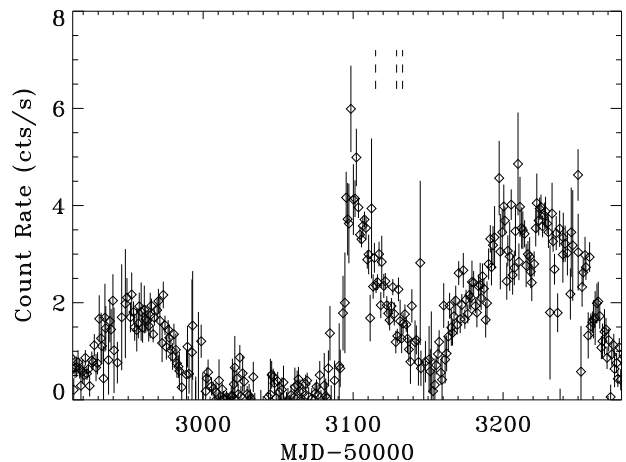,width=0.45\textwidth}
  }
  \caption{\small The {\sl Rossi}-XTE ASM (1.5-12 keV) count rate of
    LMC~X--3 plotted vs. Julian date. The short dashed bars mark the
    {\sl Chandra} observation dates. \label{fig:asm} }
\end{figure}

Our experiment used the \chandra\ low energy transmission grating plus
the high resolution camera (LETG-HRC) combination, chiefly for
its high sensitivity to detect the \ion{O}{7} K$\alpha$ line.  
The spectral resolution of the LETG-HRC is 0.05\ \AA\ (FWHM). LMC X--3 is
a variable black hole X-ray binary and its soft X-ray flux varies
quasi-periodically on a period of 99 or 199 days
\citep{cow91,cow83}. We thus proposed the experiment as a
target-of-opportunity (TOO) program during the {\sl Chandra} cycle 5.
Because of the newly imposed solar pitch angle constraints on
\chandra\ pointing duration, our proposed 100 ks exposure was split
into three separate observations of 20, 40, and 40 ksec, taken on
April 20, May 4, and May 8, 2004 (Fig.~\ref{fig:asm}).  Unfortunately,
the X-ray intensity of LMC X--3 decreased unusually rapidly from the
peak when the TOO was triggered; by the time of the latter two
(longer) observations, the count rates had dropped by a factor of
$\sim 2$.  Nevertheless, we still collected more than $4 \times 10^5$
good events in the energy range 10-40 \AA~along the two high
resolution grating arms.

We processed the LMC X-3 data, using the standard CIAO threads 
(version 3.2.1)\footnote{see http://cxc.harvard.edu/ciao.}
with the calibration database CALDB 
(version 3.0.0)\footnote{see http://cxc.harvard.edu/caldb/.}.  
As described in the Observatory Guide (version 7.0)\footnote{see http://cxc.harvard.edu/proposer/POG/html/MPOG.html}, ``the LETG diffracts X-rays into a dispersed spectrum according to the grating diffraction relation, $m\lambda=p\sin\theta,$ where $m$ is the integer order number, $\lambda$ the photon wavelength, $p$ the spatial period of the grating lines, and $\theta$ the dispersion angle.'' Therefore, events recorded at each 
$\theta$ of an observation consist of photons at multiple wavelengths
in different orders. 
Generally, this spectral order overlapping is most serious at long wavelengths,
because both the intrinsic source photon flux and 
the effective collecting area of the telescope/instrument  
decrease with photon energy. The overlapping orders cannot be sorted 
out locally in an observed spectrum, because the HRC itself lacks 
sufficient energy resolution. 

We used a forward modeling approach to globally account
for the spectral order overlapping (\S 3.1). To facilitate such modeling, we 
constructed a grating response matrix file (RMF) and an 
auxiliary response (or effective area) function (ARF) for each order
(from the 1st to the 6th) of each grating arm, by running the threads 
{\sl mkgrmf} and {\sl mkgarf}. We then added the six ARF and RMF pairs together
to form an order-combined response file 
(RSP)\footnote{see http://asc.harvard.edu/cal/Letg/Hrc\_QE/ea\_index.html\#rsps.}  for each grating arm. 
This RSP file accounts for not only the 
local line response, but also the
global spectral overlapping of the orders. For ease of processing and visualization, 
we further co-added the spectra: first the negative and positive arms, 
and then the individual observations. The corresponding RSP files 
were averaged channel by channel and weighted by the best-fit global 
models of individual spectra (see \S 3.1).  As consistency check, 
our analysis was often repeated with the individual spectra before 
the co-adding.

\subsection{{\it FUSE} Observations}

To complement the {\sl Chandra} observations, we also obtained
high-resolution ultraviolet spectroscopy of LMC X-3 via a TOO
observation with {\sl FUSE}.\footnote{For information on the design
and performance of {\sl FUSE}, see \citep{moos00, moos02}.}  The {\sl
FUSE} data cover the \ion{O}{6} resonance-line doublet at 1031.93 and
1037.62 \AA, which is valuable for understanding the ionization and
single phase vs. multi-phase nature of highly-ionized absorbers (Tripp
\& Savage 2000; Tripp et al. 2001; Mathur et al. 2003; Sembach et
al. 2004, Savage et al. 2005a,b; Collins, Shull, \& Giroux
2005). Moreover, the spectral resolution of {\it FUSE} (FWHM $\approx$
20 km s$^{-1}$) is at least an order of magnitude higher than that of any
current X-ray spectrograph. Consequently, the {\sl FUSE} observation
provides robust information about the kinematics and line broadening
as well as the likely location of the gas (the Milky Way vs. the LMC)
based on line velocities.  In CIE, the population of \ion{O}{6} ions
peaks sharply at $\sim 10^{5.5}$ K, whereas \ion{O}{7} and \ion{Ne}{9}
trace gas over a broad temperature range ($10^{5.5 -6.5}$ K;
\citet{yao05}). However, in terms of column density detection, the
{\sl FUSE} observation is more sensitive to \ion{O}{6} than the
\chandra\ observation is to \ion{O}{7} and can actually be used to
trace gas with a temperature as high as $\sim 10^{6}$ K.

The {\sl FUSE} and {\sl Chandra} observations were coordinated to
occur as simultaneously as possible. The requirement of simultaneous
X-ray and UV observations was primarily for investigation of the
accretion-disk physics and intrinsic properties of LMC X-3, which will
be presented in a separate paper.  The {\sl FUSE} data were obtained
in a single visit on 20 April 2004 for a total on-source integration
time of 88.9 ks.  The program employed the LWRS (30''$\times$30'')
aperture and time-tag mode, but only data from the LiF1 channel were
useful (other channels were not adequately co-aligned).  The data were
reduced with CALFUSE (v3.0.7) following standard
procedures.\footnote{See
http://fuse.pha.jhu.edu/analysis/analysis.html.} For {\sl FUSE}, LMC
X-3 is a relatively faint target, and it is difficult to determine
shifts between sub-exposures for alignment and co-addition. However,
since all sub-exposures were obtained during a single visit, such
shifts should be small or negligible in the LiF1 channel. The final,
co-added spectrum was binned to $\sim 7.5$ km s$^{-1}$ pixels (roughly
3 pixels per resolution element).

Before execution of our program, LMC X-3 had been observed with {\sl
FUSE} in 2002 (see Hutchings et al. 2003). While the total integration
time ($\sim 24$ ks) 
obtained by Hutchings et al. was significantly shorter than the
exposure time of the new observation, it turns out that LMC X-3 was
also substantially brighter (in the UV region near the \ion{O}{6}
doublet) when Hutchings et al. observed it. Consequently, the
signal-to-noise ratios of the old and new observations are
comparable. In principle, if some absorption is circumstellar, then
the absorption lines could vary between observations. However, we have
acquired and re-reduced the older LMC X-3 {\sl FUSE} data obtained by
Hutchings et al., and we find that the absorption lines in the old and
new data appear to be identical within the noise.  
Therefore, we have elected to coadd all of the data (LiF1a and LiF2b
from Hutchings et al. and LiF1a from the new program). Because the
flux varied significantly between the 2002 and 2004 observations, we
fitted the continua of the three spectra independently and then
coadded the continuum-normalized data, weighted inversely by their
variances. The continua were fitted with low-order polynomials in the
regions within $\pm 1000$ km s$^{-1}$ of the \ion{O}{6} $\lambda
1031.93$ line.

Hutchings et al. (2003) found that the velocity of the \ion{O}{6}
emission centroid varies by 100-150 km s$^{-1}$ as a function of the
X-ray binary phase, and they derived an orbital semiamplitude of
130-200 km s$^{-1}$ for the putative black hole in LMC X-3.  Our new
observations show a narrow \ion{O}{6} emission feature at LMC
velocities that shifts position as the binary phase changes, and the
shifts are in the same direction as reported by Hutchings et al.
However, the shifting \ion{O}{6} emission feature in the new data is
narrower than the emission presented by Hutchings et al.  We will
present a full analysis of the \ion{O}{6} emission in a separate
paper. However, it is relevant here because the narrow \ion{O}{6}
emission can lead to a complicated continuum shape against which
absorption is measured.  Misplacement of the continuum due to the
emission could cause absorption features to be incorrectly
measured. To avoid this problem, before coadding the data, we
collected the data into five phase ranges.  In each phase range, we
examined the spectrum to identify the wavelength range affected by the
narrow \ion{O}{6} emission, and we masked and excluded that wavelength
range from the final, coadded spectrum.  Because the \ion{O}{6}
emission is narrow and shifts substantially in velocity, this masking
procedure still allowed accumulation of adequate signal at all
wavelengths, i.e., no regions of the spectrum were masked out
entirely.

\section{Analysis and Results}

\subsection{{\sl Chandra} Data \label{sec:chandra}}

\begin{figure*}
  \centerline{
    \includegraphics[width=0.95\textwidth]{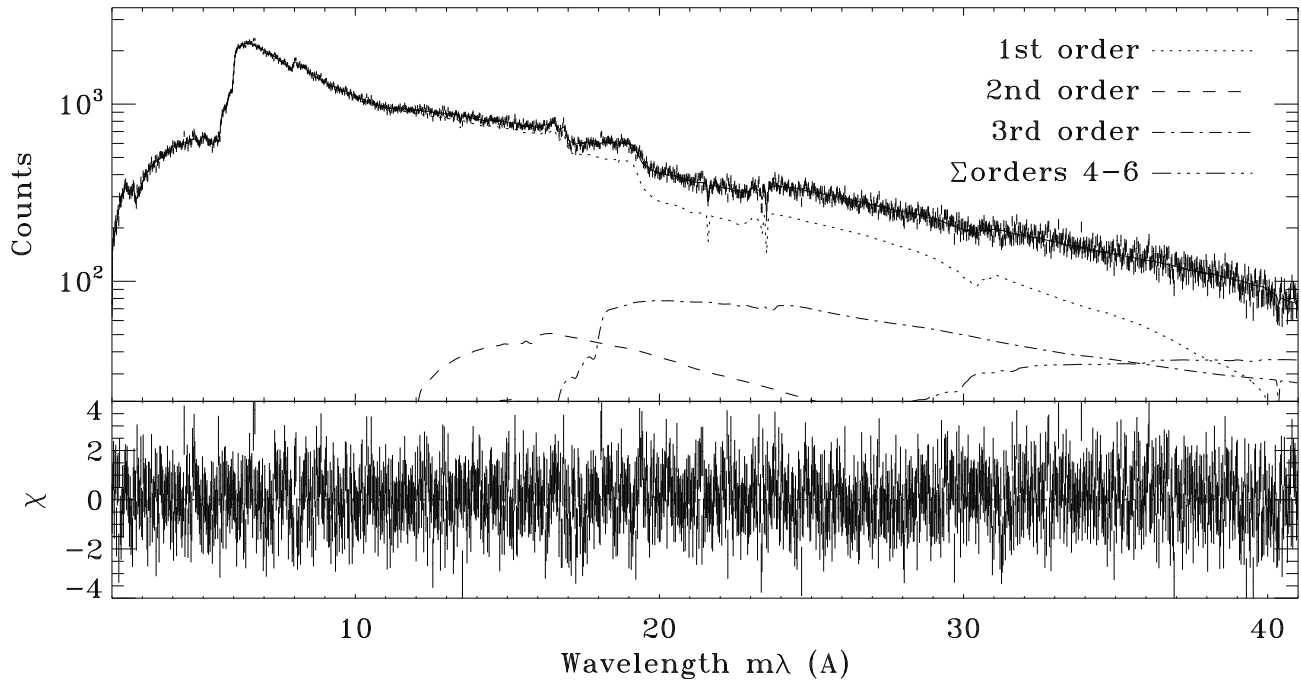}
  }
  \caption{Co-added LETG-HRC spectrum of LMC~X--3 (data points with error bars) and the best-fit continuum model (the solid curve) and the
    residuals (lower panel). While the fit is based on 
the original bin size 0.0125 \AA, this plotted spectrum has been
    grouped by a factor of 2 for clarity. 
    Contributions from individual orders are also plotted. 
    \label{fig:confusion}
  }
\end{figure*}


\begin{deluxetable}{lccc}
\tablewidth{0pt}
\tablecaption{Key Absorption Lines in the LMC X--3 
  Spectra \label{tab:lineresults}}
\tablehead{
                   & Wavelength&  $v$        & EW  \\
Line               & (\AA)     & (km s$^{-1}$) & (m\AA)}
\startdata
\ion{O}{6}              & 1031.926 & $55\pm 11$         & $235.8 \pm 37.5$ \\
\\
\ion{O}{7}  K$\alpha$   & 21.6020  & 56(-83, 194)  & 20(14, 26) \\ 
\ion{O}{7}  K$\beta$    & 18.6270  & (fixed)       & $<7.5$ \\ 
\ion{O}{8}  K$\alpha$   & 18.9670  & (fixed)       & $<4.1$ \\
\ion{Ne}{9} K$\alpha$   & 13.4470  & 112(-214, 347)& 6.3(3.7, 9.0) \\
\\
\ion{O}{1}  K$\alpha$   & 23.5140  & 128(13, 242)   & 30(23, 38)  \\
\ion{O}{2}  K$\alpha$   & 23.3492  & 283(11, 349)   & 19(12, 26)
\enddata
\tablecomments{
The errors of \ion{O}{6} measurement are at 1 $\sigma$ levels.
The rest frame wavelengths of \ion{O}{1} K$\alpha$ and \ion{O}{2} K$\alpha$ 
are taken from the averaged value in Table~\ref{tab:O7O1} 
and \citet{kaw02}, respectively.
 }
\end{deluxetable}

\begin{figure*}[thb!]
  \centerline{
    \hbox{
      \includegraphics[width=0.31\textwidth]{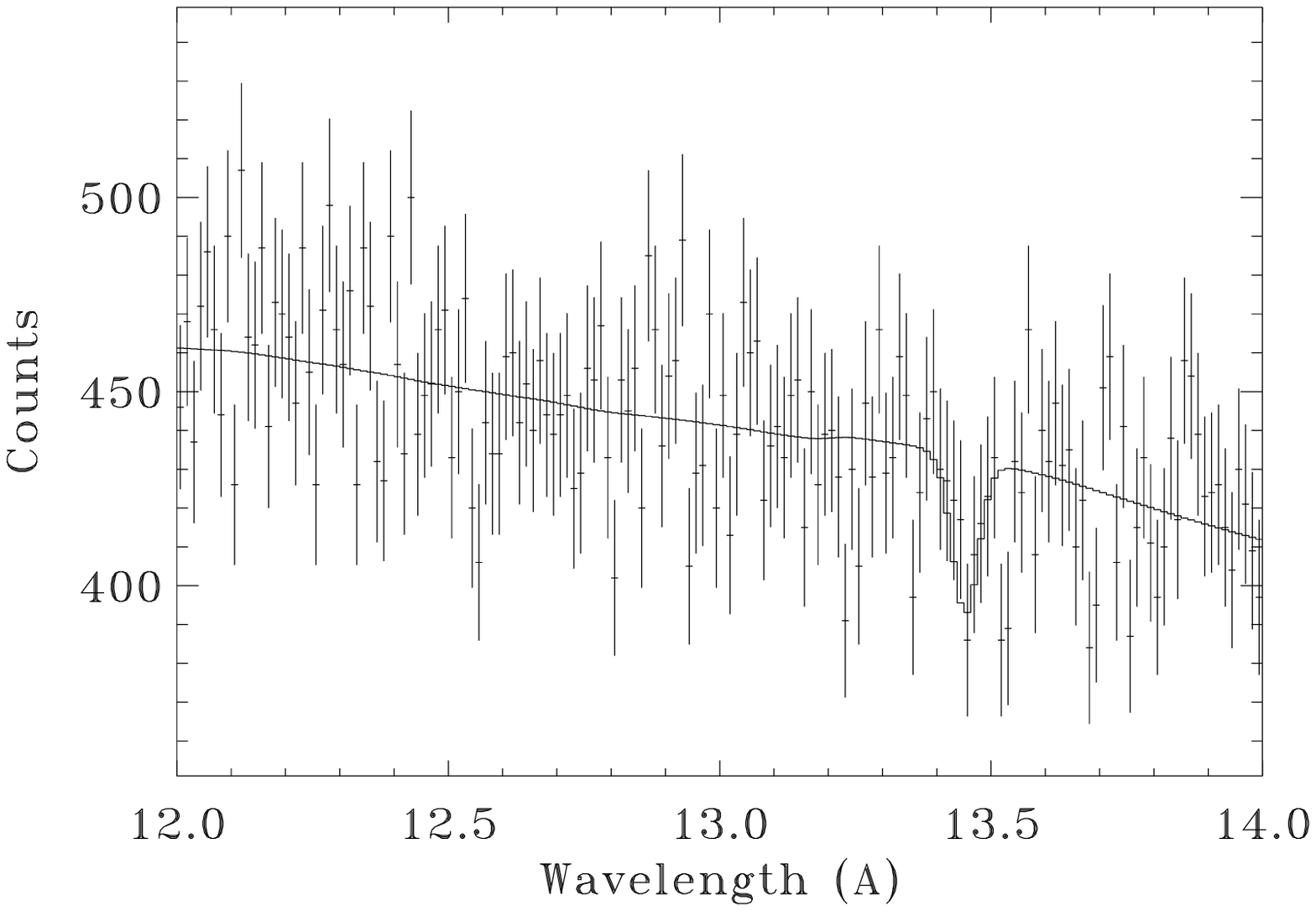}\hspace{0.05in}
      \includegraphics[width=0.31\textwidth]{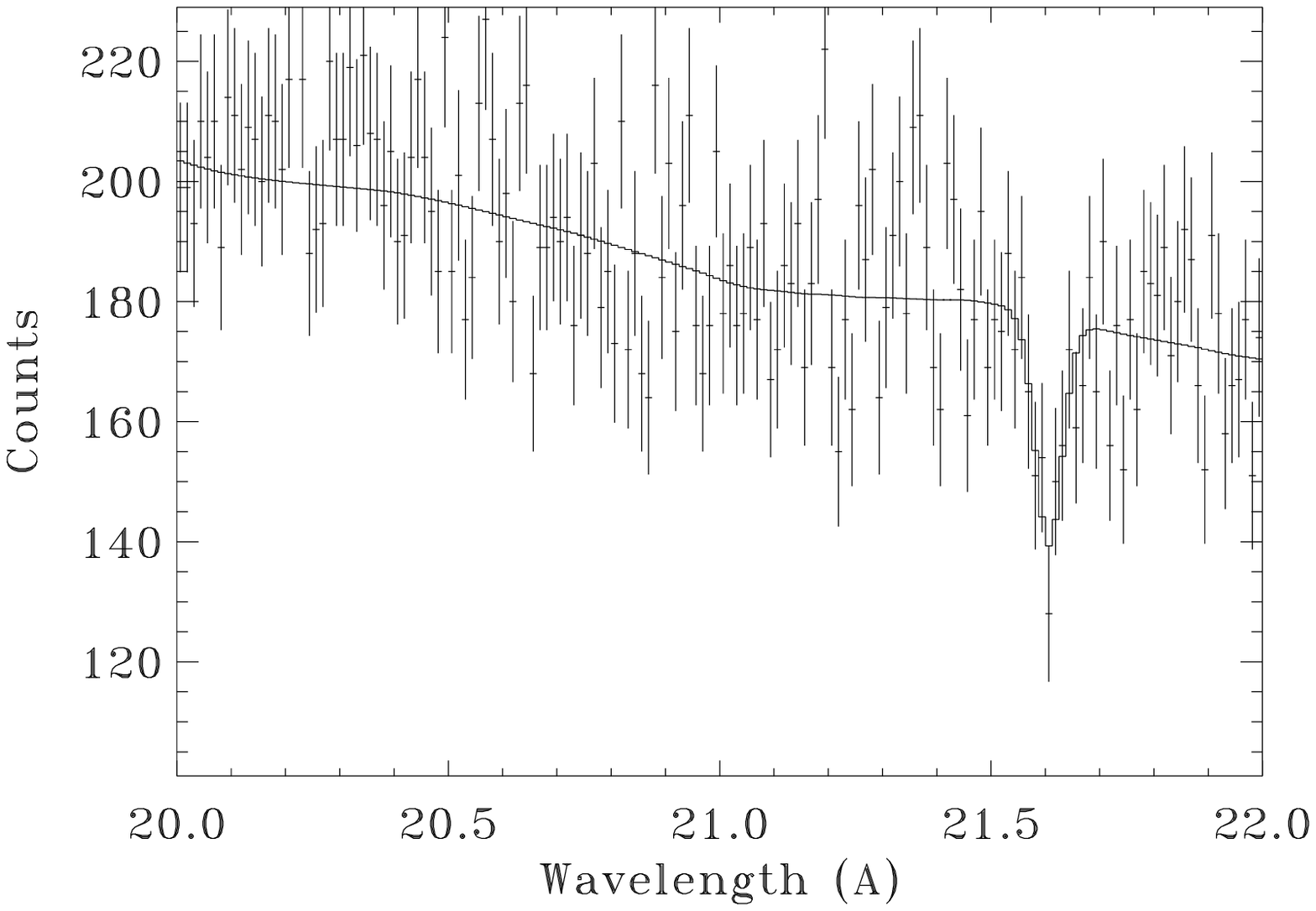}\hspace{0.05in}
      \includegraphics[width=0.31\textwidth]{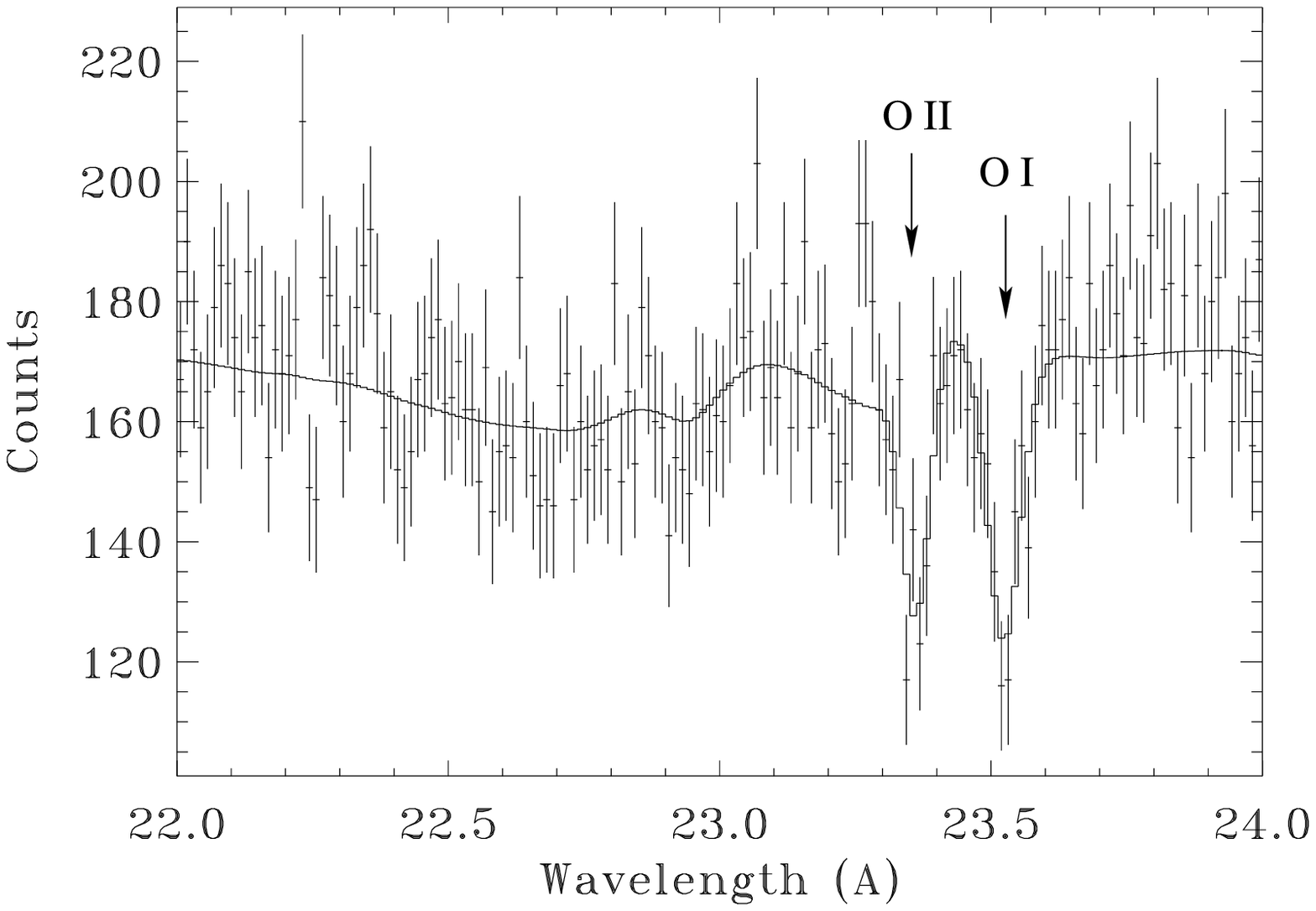}
    }
  }
  \caption{Close-ups of the {\sl Chandra} LETG-HRC spectrum of LMC~X--3 
    around the K$\alpha$ lines of (a) \neix , (b) \ovii , and 
    (c) \oi~and \oii.
    In each panel, the curves represent the fitted continuum plus 
    a negative {\sl Gaussian} line model, which has been convolved with the 
RSP.  The bin size is 0.0125 \AA.
    \label{fig:ufs}}
\end{figure*}

Our {\sl Chandra} spectral analysis uses the software XSPEC (version
11.3.1).  To properly account for the spectral order overlapping (\S
2.1), we model the entire observed spectrum in the 2-41 \AA\ range.
We first try an absorbed disk blackbody plus power-law model, but
obtain a poor spectral fit ($\chi^2/d.o.f.=3980/3114$, corresponding
to a statistical null hypothesis confidence of $\sim3\times
10^{-24}$).  This poor fit is largely due to various local deviations
from the model. These deviations are likely due to the current
imperfect calibration of the instrument response (i.e., the adopted
RSP), which does not accurately account for such effects as the
dithered HRC-S plate gaps \citep{nic05}, and to the uncertainties in
various emission and absorption models (e.g., neutral absorption
edges; Juett, Schulz, \& Chakrabarty 2004). Following \citet{nic05},
we include eight {\sl broad} Gaussian components (three positive and five
negative) to account for such local spectral deviations. The inclusion
of these broad components improves the continuum fit substantially
($\chi^2/d.o.f.=3308/3090$) and typically has insignificant effects on
{\sl narrow} lines that we want to measure.  We find that the
\ion{O}{1}, \ion{O}{2}\footnote{This line has also been identified to
be an \ion{O}{1} complex by other authors (e.g., Paerels \etal
2001).}, \ion{Ne}{9} and \ion{O}{7} K$\alpha$ absorption lines are
clearly visible at their expected rest-frame wavelengths
(Fig.~\ref{fig:ufs}; the relevant atomic parameters and references are
summarized in Table 3 of \citet{yao05}).  These lines are
characterized with four negative Gaussian profiles (each reducing the
degrees of freedom by three), which further improve the spectral fit
(reducing $\chi^2$ consecutively by 80, 24, 16, and 35).  The final
best fit is reasonably acceptable ($\chi^2/d.o.f.= 3153/3078$) and is
shown in Fig.~\ref{fig:confusion}.  The order overlapping is
considerable in the wavelength range longward of 10 \AA. The
contribution from orders higher than the 1st is $\sim$ 7, 30, 50, and
75\% at wavelength 13, 21, 33, and 40 \AA~respectively. Accounting for the
contamination from orders higher than the 1st one, we estimate the
equivalent widths (EWs) and centroid velocities (relative to
the rest frame) of the lines as presented in
Table~\ref{tab:lineresults}. We also obtain the one-sided 95\% confidence
upper bounds to
the EWs of \ion{O}{7} K$\beta$ and \ion{O}{8} K$\alpha$ absorption
lines (listed in Table~\ref{tab:lineresults}), by fixing their
centroids to the rest-frame energies and jointly-fitting their line
widths with the \ion{O}{7} K$\alpha$ line. The inclusion of these
two line components in the fit does not significantly reduce the
$\chi^2/d.o.f.$ ($= 3150/3076$). We further check the
centroid and EW of the \ion{O}{7} K$\alpha$ line in the three
observations separately, and we find no evidence for any significant
variation.

\begin{figure}
  \centerline{
      \epsfig{figure=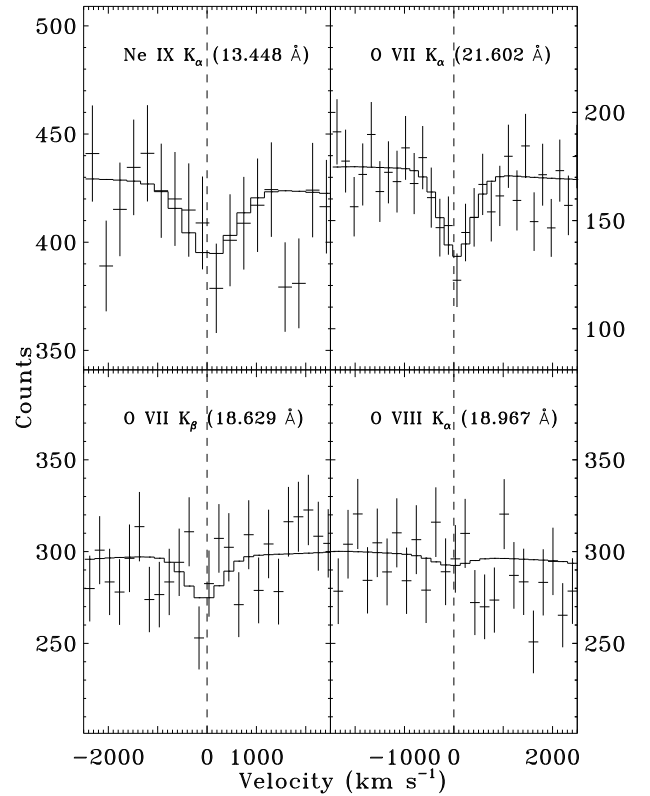,width=0.45\textwidth}
    }
  \caption{The jointly-fitted neon and oxygen absorption lines.  The
    solid histograms represent the best fit from the {\sl absline} plus
    continuum model. The
    adopted rest-frame energies (Table~\ref{tab:lineresults}) are
    marked in each panel. A negative velocity represents a blue shift.
    \label{fig:absline}}
\end{figure}

\begin{figure}
  \centerline{
    \includegraphics[width=0.45\textwidth]{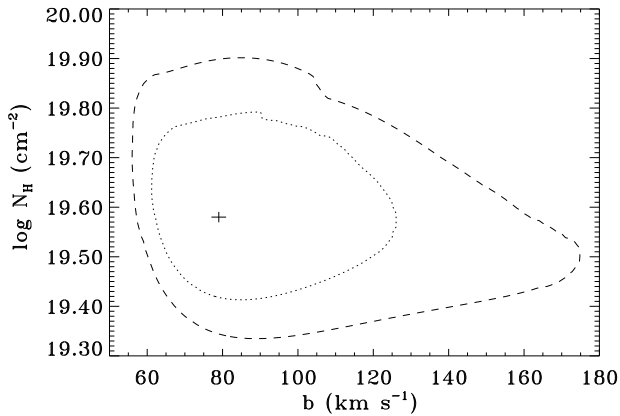}
  }
  \caption{The $N_H$--$b_v$ confidence contours at 68\% and 90\% 
    confidence levels constrained in the join-fit
    with the {\sl absline} model. \label{fig:N_b_cont} }
\end{figure}

To estimate the physical parameters of the X-ray-absorbing gas, we
jointly fit the \ion{O}{7} K$\alpha$, \ion{O}{7} K$\beta$, 
\ion{O}{8} K$\alpha$,
and \ion{Ne}{9} K$\alpha$ lines, using the multiplicative absorption line
model, {\it absline}, as detailed in \citet{yao05}. Briefly, the model
adopts an accurate absorption line profile (Voigt function) and
accounts for the line saturation automatically. This {\sl absline} fit
with three free parameters less than in the above Gaussian line fit, 
has a similar quality ($\chi^2/d.o.f.=3152/3079$).
The best-fit model is shown in Fig.~\ref{fig:absline}. The fitted
parameters are the absorbing gas temperature
log[$T({\rm K})] = 6.1(5.9, 6.3)$,
the velocity dispersion $b_v =
79(62, 132) {\rm~km~s^{-1}}$, and  the equivalent hydrogen column density
log[$N_H ({\rm cm^{-2}})]= 19.6(19.4, 19.8)$
with a corresponding 
log[$N_{\ovii} ({\rm cm^{-2}})]= 16.3(16.1, 16.5)$.
Fig.~\ref{fig:N_b_cont} shows $N_H$ -- $b_v$ confidence 
contours. The fitted line
centroid velocities are nearly identical to those in
Table~\ref{tab:lineresults}.  We can further infer the nonthermal
broadening $\xi$ from the definition of $b_v^2 = \xi^2 + 2kT/m_i$,
assuming the nonthermal motions are adequately described by a Gaussian
profile.  Here $m_i$ is the ion mass; so for oxygen, the thermal
velocity dispersion is $(2kT/m_i)^{1/2}=36(29, 47) $ km~s$^{-1}$. Therefore,
we infer the
nonthermal broadening as $\xi=70(50, 130){\rm~km~s^{-1}}$.
The {\sl absline} model fitting also allows us to predict 
log[$N_{\ovi} ({\rm cm^{-2}})]= 13.8(13.6, 14.3)$.

Because our joint fit of the absorption lines 
utilizes all the information available in the spectrum,
the source of a particular parameter constraint cannot 
be isolated easily. But foundamentally, from our measurements 
we have both lower and upper bounds on the \ovii\ and \neix\ K$\alpha$ 
absorption lines and only upper bounds on the \ovii\ K$\beta$ and
\oviii\ K$\alpha$ lines, i.e., a total six measurements. Thus we can, 
in principle, solve six limits on the parameters: $T$, $b_v$, $N_{\ovii}$, 
and $N_{\neix}$, or their combinations. With the assumption
of the Galactic ISM Ne/O ratio, we can then constrain all the parameters. 
In fact, we can do a bit better, considering that $T$ and $b_v$ are not totally
independent and that the demand for the very existence of the solution
places additional constraint on the parameters, as is demonstrated in
the following.

\begin{figure}
  \centerline{
    \includegraphics[width=0.45\textwidth]{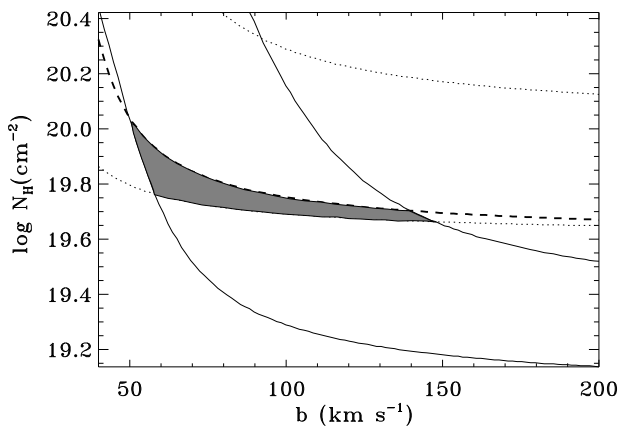}
  }
  \caption{$N_{H}$--$b_v$ diagnostic curves derived from the CoG
analysis of the \ion{O}{7} 
    K$\alpha$ ({\sl solid} curves), K$\beta$ ({\sl thick dashed}, upper bound only), 
    and \ion{Ne}{9} K$\alpha$ ({\sl dotted}) absorption lines. These curves 
    enclose the 90\% (or 95\% one-sided) confidence regions, assuming a
    CIE hot plasma with a temperature $T=1.3\times10^6$ K. 
    The overlapping region is shaded. 
    This simplistic plot is for illustration purpose only; the 
    actual measurements of the parameters are based on the joint-fits of 
    the multiple absorption lines, as detailed in the text. \label{fig:N_vs_b}
  }
\end{figure}

We can better illustrate how the specific parameter constraints
may be realized by using an alternative (but not independent) 
approach, the curve-of-growth (CoG) analysis, which is 
relatively simple and is particularly good for visualization (e.g.,
Williams \etal 2005). Fig.~\ref{fig:N_vs_b} 
illustrates how $N_H$ and $b_v$ can be constrained this way. 
For an absorption line, its EW, as measured with a negative {\sl Gaussian} model, 
is a function of $b_v$ and $N_H$, which is linked to the ionic
column density if both element abundance and plasma temperature are 
assumed. Therefore, 
from a measured EW and its uncertainties, a region can be constrained in the 
$N_{H}$--$b_v$ plot. With multiple 
absorption lines from the different transitions of the same ion 
(e.g., \ion{O}{7} K$\alpha$, K$\beta$, etc.), one may constrain
both $N_{H}$ and $b_v$ from the overlapping of the regions.
Fig.~\ref{fig:N_vs_b} illustrates that the tight constraints on both 
$N_H$ and $b_v$ in our case are derived chiefly from the 
upper and lower bounds to the \ion{O}{7} K$\alpha$, 
the lower bound to \ion{Ne}{9} K$\alpha$, and the upper bound to 
the \ion{O}{7} K$\beta$ EW measurements. Apparently, these four
EW bounds do not necessarily lead to a non-zero overlapping region,
depending on the assumed $T$ and Ne/O ratio. 
Therefore, the four EW bounds give constraints not only on both  $N_H$ and 
$b_v$, but also on the combination of the temperature and abundance, 
though indirectly. Of course, 
the upper bounds to \ion{O}{7} K$\beta$ and \ion{Ne}{9} K$\alpha$ 
provide additional constraints, especially on the upper limit to $T$. 
This apparent over-constraining power of the EW 
measurements arises from the demand for the existence of a solution 
(an overlapping region in a multi-dimensional parameter space).

The CoG approach, however, has its
drawbacks. First, it does not use all the information in the spectrum, at least
not in a coherent fashion; absorption lines are measured independently (thus 
the improved counting statistics from a combination of multiple weak lines is 
not capitalized). Second,
the joint-analysis of two or more lines in a CoG diagnostic
plot, though
informative, does not allow for a rigorous error propagation of the parameter
constraints; for example, it is difficult to quantify the overlapping 
significance and/or to include the uncertainties in the assumed temperature  
(Fig.~\ref{fig:N_vs_b}). 
Third, the EW measurement from the negative {\sl Gaussian} model
can be problematic, especially when the absorption line is saturated
(e.g., an optical depth $\sim 5$ at the line center of the \ion{O}{7} 
K$\alpha$ line in the spectrum of LMC~X--3); the EW value
can be significantly over-estimated (see Fig. 4 in Yao \& Wang 2005). 
In fact, a plot like Fig.~\ref{fig:N_vs_b} represents only a cut in a
multiple dimensional parameter space constrained by the absorption line
measurements (e.g., via an {\sl absline} model fit). 
Therefore, we use the CoG approach only for the visualization 
and consistency check of the absorption line measurements.

\subsection{\fuse\ Data}

\begin{figure}
  \centerline{
    \epsfig{figure=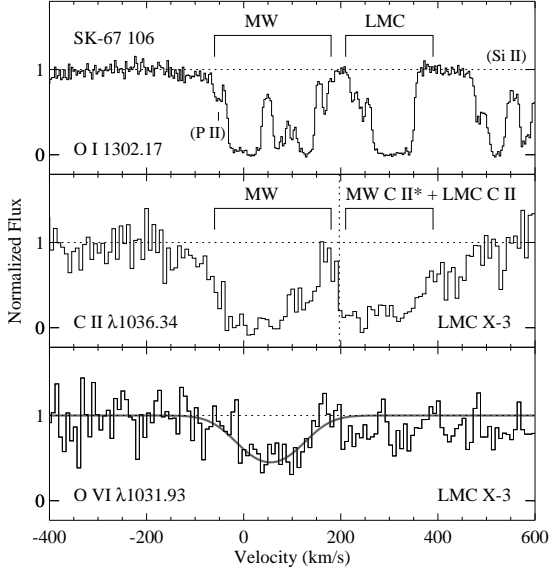,width=0.45\textwidth}
  }
  \caption{\small Continuum-normalized UV absorption line profiles
  recorded in the spectrum of LMC X-3, including (middle) {\sc C ii}
  $\lambda$1036.34 and (bottom) \ion{O}{6} $\lambda$1031.93, plotted
  versus heliocentric velocity.  For comparison, the \ion{O}{1} 
  $\lambda$1302.17 observed with the STIS E140M echelle toward the B
  star SK-67 106 in the LMC is plotted in the top panel. The vertical
  dotted line in the middle panel indicates the location of
  \ion{C}{2}* $\lambda$1037.02 absorption at $v = 0$ km s$^{-1}$.  A
  \ion{P}{2} line is present near the \ion{O}{1} absorption as
  indicated, but this very weak transition only affects the profile at
  the marked velocity. The strong \ion{Si}{2} $\lambda$1304.37 line is
  present at $v >$ 450 km s$^{-1}$ in the top panel, but this does not
  affect the \ion{O}{1} absorption at $v < 400$ km
  s$^{-1}$. Note that the regions used for continuum fitting extend beyond the
velocity limits of this plot.
\label{fig:fusespec} }
\end{figure}

We analyzed the {\it FUSE} observation entirely independently from the
{\it Chandra} results, i.e., no constraints from the X-ray data were
imposed on the {\it FUSE} measurements.  The low UV flux level makes
LMC X-3 a challenging target for {\it FUSE}, and most of the useful
information in the {\it FUSE} spectrum is in the 1030-1040 \AA\ range
where the flux is elevated by a broad emission feature \citep{Hut03}. The
middle and lower panels in Fig.~\ref{fig:fusespec} show the
interstellar \ion{C}{2} $\lambda$1036.34 and \ion{O}{6}
$\lambda$1031.93 lines recorded in the coadded {\it FUSE} LMC X-3
spectrum. For purposes of comparison, we also show in the upper panel
in Fig.~\ref{fig:fusespec} the interstellar \ion{O}{1}
$\lambda$1302.18 absorption observed with the Space Telescope Imaging
Spectrograph (STIS) toward SK-67 106, a hot star located in the main
body of the LMC (see Fig.~\ref{fig:lmc_ha}).\footnote{The STIS
observations of SK-67 106 were obtained with the E140M echelle mode
(spectral resolution = 7 km s$^{-1}$ FWHM) and reduced as described in
\citet{tripp01}.  While it might ostensibly seem more sensible to
present the \ion{C}{2} absorption observed toward SK-67 106 since
\ion{C}{2} is detected toward LMC X-3, we prefer the \ion{O}{1} line
for two reasons: first, toward SK-67 106, \ion{C}{2} is badly
saturated and all of the component structure evident in
Fig.~\ref{fig:fusespec} is hidden in the black, saturated cores of the
\ion{C}{2} lines; second, and more importantly, \ion{C}{2} at LMC
velocities is badly confused by blending with \ion{C}{2}* at Milky Way
velocities (see text).}

The SK-67 106 panel in Fig.~\ref{fig:fusespec} shows that in the
general direction of the LMC, strong and complex Milky Way absorption
is detected at $0 \lesssim v \lesssim 200$ km s$^{-1}$, and strong
(saturated) absorption due to the ISM of the LMC is present at $200
\lesssim v \lesssim 400$ km s$^{-1}$.  Toward stars located in the
main body of the LMC, strong \ion{O}{6} absorption is frequently
detected within these velocity ranges as well (e.g., Howk \etal 2002;
Danforth \etal 2002). Toward LMC X-3, \ion{C}{2} and \ion{O}{6}
absorption lines are clearly detected in the Milky Way velocity
range. While a strong absorption feature is evident in the LMC X-3
spectrum at the velocity expected for \ion{C}{2} in the LMC ISM, much
of that absorption is likely due to a blend of \ion{C}{2}*
$\lambda$1037.02 and \ion{O}{6} $\lambda$1037.62 from the Milky Way,
and it is difficult to reliably assess the amount of LMC \ion{C}{2}
(or even if it is present at all) in the direction of LMC
X-3.

We have measured the equivalent width and column density of the
\ion{O}{6} $\lambda$1031.93 absorption by direct integration using the
techniques of Sembach \& Savage (1992), including assessment of the
continuum placement uncertainty as well as statistical uncertainty in
the overall error bars. From inspection of Fig.~\ref{fig:fusespec},
we see that \ion{O}{6} absorption is clearly detected at Milky Way
velocities ($v \lesssim 200$ km s$^{-1}$). The \ion{O}{6} EW
integrated over the Milky Way velocity range is listed in
Table~\ref{tab:lineresults}; the EW indicates that the Galactic
\ion{O}{6} absorption is detected at 6.3$\sigma$ significance.  The
\ion{O}{6} absorption at LMC velocities is much weaker, and
integrating over the LMC velocity range ($190-390 {\rm~km~s^{-1}}$), 
we obtain EW(LMC) =
96.2$\pm$46.4 m\AA , i.e., only a 2.1$\sigma$ feature. Moreover, if we
compare the three individual spectra (LiF1a and LiF2b from Hutchings
et al. and LiF1a from the new observation) that were coadded in
Fig.~\ref{fig:fusespec}, we find that Milky Way \ion{O}{6} is
detected at $> 3\sigma$ in all three individual observations, and the
EWs measured from the individual spectra agree within their 1$\sigma$
uncertainties. However, \ion{O}{6} absorption at LMC velocities is not
significantly detected in any of the individual spectra. We conclude
that the current data do not show compelling evidence of \ion{O}{6}
absorption at LMC velocities. While observations of stars in the main
body of the LMC almost always show \ion{O}{6} absorption associated
with the LMC (Howk et al. 2002), LMC X-3 is outside of the main body
of the LMC (see Fig. 1), and therefore it is not necessarily
surprising that the \ion{O}{6} is weaker in this direction. If we were
to include the 2$\sigma$ absorption in the LMC velocity range, this
would increase the total column density by $\approx$ 40\% .

Over the Milky Way velocity range, we find 
log[$N_{\ovi} ({\rm cm^{-2}})]= 14.42_{-0.09}^{+0.07}$ (1$\sigma$
uncertainties) from direct integration. Assuming a single
Gaussian component for \ion{O}{6}, we have also used the Voigt profile
fitting software of Fitzpatrick \& Spitzer (1997) to measure the
\ion{O}{6} centroid, line width, and column density, and we obtain the
fit shown in Fig.~\ref{fig:fusespec} with $v$(\ion{O}{6}) $= 55\pm
11 {\rm~km~s^{-1}}$, $b_v$(\ion{O}{6}) $= 81^{+18}_{-15}
{\rm~km~s^{-1}}$, and 
log[$N_{\ovi} ({\rm cm^{-2}})]= 14.50\pm0.07$. The \ion{O}{6} columns obtained from direct
integration and profile fitting are in good agreement. The \ion{O}{6} column
predicted from the best-fit
{\sl absline} model of the X-ray absorption lines (\S 3.1) is a factor of
$\sim5$ lower; but because of the large statistical uncertainties 
in the X-ray data,  the discrepancy is only at the $\sim 2 \sigma$ 
confidence level.  
Interestingly, the \ion{O}{6}
$v$ and $b_v$ values are consistent with those of \ion{O}{7}
(see Table~\ref{tab:lineresults} and \S \ref{sec:chandra}).

\section{Discussion}

\subsection{Origin of the far-UV and X-ray absorbers}

The centroid positions of the observed absorption lines (Table 1) may
be used to determine their origin, Galactic or the LMC. Our measured
centroid position of the \ion{O}{7} K$\alpha$ line (Table 1) is
consistent with the rest-frame energy and is significantly {\sl
inconsistent} with an LMC origin. The \ion{O}{6}  line centroid differs by
$>10 \sigma$ from LMC velocities. Therefore, we can conclude with
great confidence that the bulk of the \ion{O}{6} absorption detected
toward LMC X-3 does not originate in the LMC.  However, it remains
possible that the \ion{O}{6} and \ion{O}{7} absorption occurs in
separate, unrelated gas clouds, so it is important to
scrutinize the disagreement between the \ion{O}{7} centroid and the
LMC velocity.  The question is whether or not any systematics may
account for this disagreement.  We address this question, both
empirically and physically.
\begin{deluxetable}{lccc}
\tablewidth{0pt}
\tablecaption{Comparison of line centroid positions\label{tab:O7O1}}
\tablehead{
                        & \ion{O}{7}                   & \ion{O}{1}       & $\Delta \lambda$  \\
  Sources (OB ID)       & (\AA)                  & (\AA)    & (\AA)   }
\startdata
  X1820--303 (98)         & $21.614\pm9$        &  $23.508\pm9$ & 1.894$\pm13$\\
  MRK 421 (4149)          & 21.603$_{-3}^{+9}$   &  $23.520\pm9$ & $1.917^{+13}_{-9}$ \\
  PKS 2155--304 (331)     & $21.615_{-14}^{+24}$ &  23.514$_{-23}^{+21}$ & $1.899_{-27}^{+32}$\\
  \hline
  Mean of the above       & 21.608$_{-3}^{+6}$   & 23.514$\pm6$    & 1.906$_{-7}^{+9}$\\
  \hline
  LMC X--3                & $21.606\pm10$ & $23.524\pm9$ & $1.918\pm13$
     
\enddata
\tablecomments{
The 90\% confidence error bars are in units of m\AA. }
\end{deluxetable}

Although individual lines are not resolved in the \chandra\
observations, we may determine the line centroids to an accuracy
(depending on the counting statistics) comparable to the absolute
photon energy scale uncertainty of the instrument.  We adopt that the
uncertainty to be $\sim 9$ m\AA~(0.19 eV) for the
LETG-HRC\footnote{see the Chandra calibration website for CIAO 3.2
with CALDB 3.0.1; http://asc.harvard.edu/cal/}, although higher values
(as high as $\sim 50$ m\AA) are also cited in documents\footnote{see
http://cxc.harvard.edu/proposer/POG/html/MPOG.html}, probably
resulting from evolving calibration accuracy at various stages.
Table~\ref{tab:O7O1} compares the line centroid measurements of LMC
X--3 with three other sources that have good LETG-HRC detections of
both \ion{O}{1} and \ion{O}{7} absorption lines. While X1820--303 is a
Galactic LMXB \citep{fut04, yao05}, the other two sources are AGNs
(Williams \etal 2005; Yao et al. 2005, in preparation).  The
\ion{O}{7} line centroid measurements are all consistent within $\sim
1 \sigma$ (Table~\ref{tab:O7O1}). A maximum likelihood analysis
indicates no significant intrinsic (non-statistical) dispersion in
these measurements, but the allowed 1$\sigma$ upper limit is
consistent with the adopted absolute photon energy scale
uncertainty. Accounting for both the statistical and systematic
uncertainties, we conclude that the LMC origin of the highly-ionized
X-ray absorption lines detected in the LMC X--3 observations can be
ruled out kinematically at $\sim 2\sigma$ confidence.

What is the origin of the cool oxygen absorption lines? 
\oi\ and \oii\ trace the cold and warm phases of the ISM \citep{field71}. 
From the EW measurements of the \oi\ and \oii\ K$\alpha$ absorption lines,
the CoG plots of \citet{Juett04}, and the Galactic ISM oxygen abundance,
we estimate the equivalent hydrogen column densities of the cold and warm
 phases as 3.2(1.9, 5.3) $\times 10^{20} {\rm~cm^{-2}}$ and
1.2(0.5, 2.0) $\times 10^{20} {\rm~cm^{-2}}$, respectively (neglecting 
a potenital, but typically small \oiii\
contribution). These column densities, especially their
lower limits, are not sensitive to the
exact $b_v$ (assumed to be $20 {\rm~km~s^{-1}}$; \citet{Rey88}) and
are consistent with $3.8^{+0.8}_{-0.7} \times
10^{20} {\rm~cm^{-2}}$ estimated from the oxygen absorption edge 
analysis of an {\sl XMM-Newton} grating spectrum of LMC X--3 \citep{page03}.  
The consistent column density values are also
obtained in modeling the broad band
continuum observed with {\sl BeppoSAX} \citep{yao05b}
and \chandra\ (\S~1). The warm gas column density is in good 
agreement with the value expected from the so-called Reynold's layer
of our Galaxy \citep{Rey89}. The cold gas is further 
traced by the 21 cm emission 
line (though with a poor spatial resolution), which gives a neutral hydrogen 
column density of $\sim 4.6 \times 10^{20} {\rm~cm^{-2}}$ in the LMC X--3 field
and shows a mean velocity of $\sim 10 {\rm~km~s^{-1}}$
\citep{Stave03, Dickey90}. There is little emission in the 100-430
${\rm~km~s^{-1}}$ velocity range ($\lesssim 5 \times 10^{19}\ \rm
cm^{-2}$). These values are consistent with the \ion{C}{2} absorption seen in
the UV spectrum (Fig.~\ref{fig:fusespec}); bearing in mind that a
substantial portion of the \ion{C}{2} feature at LMC velocities is
actually due to \ion{C}{2}* and \ion{O}{6} $\lambda$1037.62 at $v = 0$
km s$^{-1}$, any remaining \ion{C}{2} at $v \approx$ 300 km s$^{-1}$
could easily arise in gas with $N$(\ion{H}{1}) $\ll 10^{19} {\rm~cm^{-2}}$. 
Therefore, the cool oxygen absorption lines are also primarily Galactic. 

The Galactic origin of the \ion{O}{1} and \ion{O}{2} absorption lines is
{\sl not} inconsistent with their velocity centroid listed in 
Table~\ref{tab:lineresults}.
There are considerable wavelength uncertainties in both
the theoretical calculations and the laboratory measurements of the 
\ion{O}{1} and \ion{O}{2} lines. Laboratory measurements of the \ion{O}{1}
K$\alpha$ transition, for example, range from 23.489 to 23.536 \AA\ 
\citep{stol97}, which is at least partially due to systematic uncertainties 
in the absolute wavelength calibration, and are probably less accurate than 
X-ray absorption line observations \citep{jue05}.  
The mean centroid of the \ion{O}{1} K$\alpha$ lines listed in
Table~\ref{tab:O7O1} is between 
the LETG-HRC measurement of LMC X--3 and the mean value obtained by
\citet{Juett04} from a similar analysis of \chandra\ MEG-ACIS
observations of Galactic LMXBs.  All these line centroid measurements
agree with each other within $ 2 \sigma$ statistical uncertainties. 
It is based on these considerations that we decided to use the mean \ion{O}{1} 
K$\alpha$ line centroid in Table~\ref{tab:O7O1} as a reference for the 
velocity shift in Table~\ref{tab:lineresults}. Furthermore, LMC X--3 has 
the Galactic coordinates of $l,b = 273\fdg576,-32\fdg082$;
the conversion of our measured line-of-sight velocities into values in
the local standard of rest or in the Galactic standard of rest
requires a correction of -13 or -199 ${\rm~km~s^{-1}}$, respectively (assuming
a Galactic rotation speed of 220 ${\rm~km~s^{-1}}$ in the solar
neighborhood and neglecting the smaller barycenter correction). 
Therefore, the \ion{O}{1} and \ion{O}{2} K$\alpha$  line centroids,
though with large uncertainties, are consistent with an origin
in the local ISM.

We may now further constrain the origin of the \ion{O}{7} absorption line
from its energy separation from the \ion{O}{1} line. This relative line
separation should be independent of the absolute energy scale accuracy
of the observations.  Table~\ref{tab:O7O1} shows that the separation
$1918\pm13$ m\AA~for the LMC X--3 is consistent with the mean
value ($1906_{-7}^{+9}$ m\AA) for the other sources and is 
grossly inconsistent with the value (1882 m\AA) if \ion{O}{7} is 
redshifted to the LMC X--3 velocity. Therefore, the \ion{O}{7} absorption
line most likley does not originate in the LMC.

Of course, the above consideration of the line positions alone cannot
be conclusive about the exact origin of the \ion{O}{7} absorption line.
The systemic redshift of LMC X--3 might be compensated by a blueshift of
the \ion{O}{7} absorber.  LMC X--3 is an X-ray binary of an unseen primary
compact object, which has been interpreted as a 10-15 $M_\odot$ black
hole, and a secondary star (B3 V) with an orbital period of 1.7 days
\citep{cow83, Hut03}. The X-ray emission is apparently powered by
accretion via Roche lobe overflow (\citet{Nowak01} and references
therein). An absorption line
produced by a potential accretion disk and/or stellar wind should be
broad and blue shifted. The expected velocity width and shift should be
comparable to the escape velocity, on the order
of $10^3 {\rm~km~s^{-1}}$ (e.g., \citet{yao05, pae00}), which is not
consistent with our measurement of the \ion{O}{7} line (Table 1).
Similarly, one may expect to detect a strong \ion{O}{6} absorption by 
a wind, probably in its outer regions where the ionization parameter becomes 
relatively small.
Clearly, such a wind scenario  is {\sl inconsistent} with the constancy, 
narrowness, and rest-frame velocity of the \ion{O}{6} absorption 
detected with {\sl FUSE}. In addition, there is no sign for any proximity effect of LMC 
X--3, an optical nebula or significant absorption by photo-ionized species 
(e.g., \citet{yao05}), indicating a lack of
gas around the X-ray source.
We thus tenatively conclude that the \ion{O}{7} as well as 
the \ion{O}{6} and \ion{O}{1}
absorption lines arise in the gas in and/or around the Galaxy.

\subsection{Properties of the far-UV and X-ray-absorbers \label{sec:abund}}

Clearly, the inference of the total column density of the X-ray-absorbing
gas depends on its absolute metal abundance, which is yet to be determined
observationally. We may
use the dispersion measures (DMs) of pulsars observed in the vicinity
of LMC X-3 to estimate the total number of free electrons along the
sight-line. Among the four known radio pulsars in the LMC, PSR J0502-6617
and J0529-6652 are close to the LMC X-3 sight-line
(Fig.~\ref{fig:lmc_ha}) and have the DMs of 68.9($\pm3$) and 103.2($\pm3$) 
${\rm~pc~cm^{-3}}$ (1$\sigma$ error bars; \citet{craw01}). 
The difference in the DMs presumably
reflects the different contributions of free electrons in the LMC,
because both pulsars are projected in regions with significant local
H$\alpha$ emission (Fig.~\ref{fig:lmc_ha}). In addition, the Galactic warm
\ion{H}{2} medium is also expected to be a main (probably dominant)
source of the free electrons.  Therefore, we can only get an upper
limit to the total hot gas column density from the DMs. The lower DM
value of J0502-6617 ($l,b = 273\fdg576,-32\fdg082$), presumably least contaminated by the  warm \ion{H}{2}
medium, is only slightly higher (by a factor of 1.5) than that
expected from Galactic pulsar DMs, but is almost same as its highest
individual value (accounting for the Galactic latitude dependence,
i.e., sin~$b$). From the comparison of our $N_{\ovii}$ estimate
(\S 3.1) with the J0502-6617 DM value, we obtain an
oxygen abundance as $\sim 19(12-31)\%\xi^{-1}$ of 
the Galactic ISM value (\S 1),
where $\xi$ is the fraction of the electrons in the hot gas. If our 
estimate of the warm \ion{H}{2} column density from the \oii\  
measurement is reasonable (\S 4.1), i.e., $1-\xi \approx 0.57(0.24-0.95)$, 
the oxygen abundance can then be close to the ISM value or even substantially 
higher. Although various relevant systematics are difficult
to quantify here, the lower limit to the oxygen abundance, 
16\% (or an O/H number ratio of $8 \times 10^{-5}$), should be quite firm.
Therefore, the hot gas is unlikely to be
accreted recently from the intergalactic medium (IGM).

\begin{figure}[thb!]
  \centerline{
    \epsfig{figure=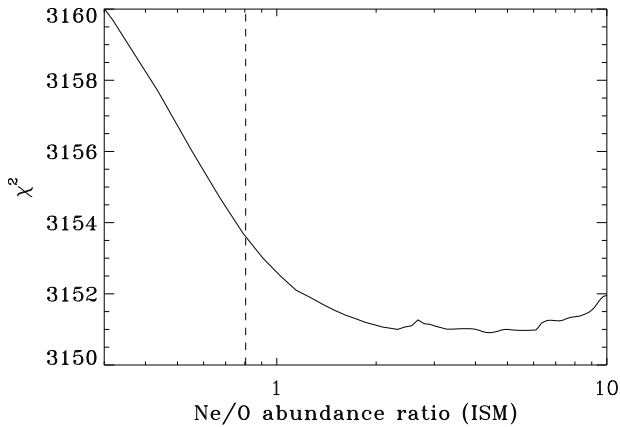,width=0.45\textwidth}
  }
  \caption{\small $\chi^2$ as a function of the Ne/O  ratio in a {\sl absline} fit to the X-ray absorption lines. The vertical dashed line marks the 95\%
confidence lower limit to the  Ne/O  ratio. 
\label{fig:chi_ne_o} }
\end{figure}

We may also constrain the Ne/O abundance 
ratio, which is used in the inference of the hot gas parameters. 
Published X-ray studies are all based on  measurements of absorption edges, 
which are produced 
primarily by {\sl cool} gas. The neutral Ne/O ratio has been determined to be 
$1.67\pm0.11$ and $1.10\pm0.36$ of the assumed ISM value in the sight-lines of 
the X-ray binaries Cygnus~X--1 \citep{sch02b} and X--2 \citep{tak02}, 
respectively. Several X-ray binaries indicate unusually
high Ne/O ratios in X-ray-absorbing gas (a factor of a few higher; 
\citet{pae01,jue05}).  Such high Ne/O ratios,
very difficult to explain in the ISM context, are believed to arise
from material local to the binaries, which typically show either
coronal emission lines or burst behavior. In one occasion, the Ne/O
ratio apparently varies with time \citep{jue05}. With the measurements
of the line absorption by the highly-ionized species, it is now also possible
to measure the relative metal abundances of {\sl hot} gas. While we are not 
aware of such a measurement in publication, a joint analysis of
multiple absorption lines observed in the spectra of the X-ray source 
4U1820-303 without apparent local contamination does
indicate a normal Ne/O ratio of the X-ray-absorbing hot gas 
(Yao \& Wang, in preparation). Similarly, if the
ratio is allowed to freely vary in our {\sl absline} fit of the LMC X-3 
X-ray spectrum in \S 3.1, we obtain  a 95\%
confidence lower limit of 0.8 times the ISM value, or a Ne/O elemental
number ratio of 0.14; the upper limit cannot be meaningfully constrained 
(Fig.~\ref{fig:chi_ne_o}). Therefore, 
the Ne/O ratio of the hot gas along the LMC X-3 sight-line is 
consistent with a normal ISM value.

In the {\sl absline} fit, the hot gas temperature decreases with the increase 
of the assumed Ne/O ratio value. But $N$(\ion{O}{7}) (or $N_H$) hardly 
changes, because the ionic fraction ($\sim 1$) 
is insensitive to the exact value of the temperature in the range of
$10^{5.5}-10^{6.2}$ K (e.g., Yao \& Wang 2004). Similarly, $b_v$, 
especially its most useful lower limit, is insensitive to the Ne/O ratio value
(Fig.~\ref{fig:N_vs_b}).

How are the absorbers as observed in the LMC X-3 sight-line 
related to the general hot ISM in the Galaxy?
 The two AGN sight-lines nearest to LMC X-3 (NGC~1705 and PKS
 0558-504) do show low-velocity \ion{O}{6} absorption features (239
 and 257 m\AA\ respectively, with statistical errors of about 17 m\AA,
 see Wakker et al. 2003), consistent with our above measurement for
 the LMC X-3 sight-line.  The inferred physical properties of the
 X-ray-absorbing gas along this sight-line are also comparable to
 those along the sight-lines away from the diffuse soft X-ray emission
 enhancement in the Galactic central region (Snowden et al. 1997).
 Based on nine similar detections (mainly \ion{O}{7} K$\alpha$ and Ne
 IX K$\alpha$ absorption lines) in the spectra of Galactic X-ray binaries
and AGNs, 
Yao \& Wang (2005) show that the hot
 gas column density along the sight-line to LMC X--3 is consistent
 with a global characterization of the hot ISM in the Galaxy. Assuming
 a disk morphology, the vertical exponential scale height of the hot
 gas is $\sim 1.2$ kpc. Compared with the previously measured scale
 heights of $\sim 2.7, 3.9, 4.4$, and $5.1$ kpc for \ion{O}{6},
 \ion{N}{5}, \ion{C}{4}, and \ion{Si}{4}, respectively (e.g., Savage
 \etal 1997, 2000), the hot gas appears to follow a trend from high to
 low ionization states away from the Galactic plane [i.e., the scale
 height apparently {\it increases} with the decreasing ionization
 potential of the species \citep{wang05}]. Such an ionization
 stratification may be understood as a result of a Galactic fountain,
 in which hot gas heated primarily by supernova blastwaves in the
 Galactic disk cools off on the way to the Galactic halo \citep{Sha76,
 bre80}.

A Galactic fountain scenario may also explain the nonthermal 
broadening of the far-UV- and X-ray-absorbing gas (\S 3).  
After the correction for the local standard of rest, the 
far-UV- and X-ray-absorbing hot gas along the sight-line to 
LMC X--3 seems to have a systemic velocity close to zero,
relative to the cold gas in the solar neighborhood (Table 1). 
Therefore, the hot gas should 
nearly co-rotate with the stellar
(or cool gas) disk. Because of the high Galactic latitude location
of LMC X--3, the line-of-sight
velocity of the hot gas should hardly change through the disk with 
a flat rotation curve in the solar neighborhood. The expected lagging
``halo'' effect, as observed in diffuse cool and warm gases 
around nearby edge-on disk galaxies (Collins
\etal 2002 and references therein), should be small on scales
of the hot gas disk. Therefore, the presence of the nonthermal
broadening may suggest a significant vertical velocity 
gradient of the hot gas, as is expected in a Galactic fountain
\citep{Sha76, bre80}.

In this Galactic fountain scenario, and in the general ISM context,
\ion{O}{6} and \ion{O}{7} are expected to co-exist globally, though 
probably not in the exactly same phase. 
As an extreme case, we may assume that the observed 
\ion{O}{6} column all arises in the
X-ray-absorbing gas and may then jointly fit the X-ray absorption 
lines with the \fuse\
\ion{O}{6} line. This fit gives log$N_H$, $b_v$, log$T$, and
the Ne/O ratio of the absorbing gas as 19.5(19.3, 19.8), 74(61, 94),
5.82(5.75, 5.91), and 3.1($> 1.3$), respectively. The temperature
obtained here represents the lowest possible limit; otherwise,
the observed \ion{O}{6} absorption should be stronger.  
If only a fraction
of the observed \ion{O}{6} absorption arises in the X-ray-absorbing gas,
the temperature (Ne/O ratio) would then be greater (smaller). 
In general, one needs to consider the possible multi-phase nature 
of the hot gas. \ion{O}{6} traces hot gas at the peak of its 
cooling curve and thus typically represents a very unstable transient phase, 
as is expected in the cooling hot gas of the Galactic fountain and/or
 at interfaces between hot and cool gases. Of course, the gas could 
in principle be in a non-CIE state, although its effect does not seem to
be important for 
such highly-ionized species as \ion{O}{6} and \ion{O}{7} in a typical
ISM cooling flow \citep{Sutl97}. While a comprehensive examination of these 
potential complications requires data of better quality, the results 
from the above far-UV and
X-ray joint-fit should be used with extreme caution. 

\subsection{Implication for other $cz \sim 0$ \ion{O}{7} absorbers}

The above discussion corroborates the
conclusion that the bulk of the highly-ionized gas is located
near the Milky Way, most likely in a thick hot gaseous disk. 
The presence of this disk has strong implications for 
the interpretation of similar $cz \sim 0$ \ion{O}{7} absorbers
observed in AGN spectra. Both the
temperature and the line broadening of the hot gas in the
LMC X--3 sight-line are comparable to those of the absorbers in the
AGN sight-lines. In the sight-line to Mkn 421 ($l, b =179\fdg83,
65\fdg03$) in particular, which is also away from the Galactic central
X-ray emission enhancement, the
inferred vertical gas column density $N_H |{\rm sin}~b|\approx 2.1
\times 10^{19} {\rm~cm^{-2}}$  (Yao \& Wang 2005) 
agrees well with our measurement of
$2.1(1.3, 3.4)\times 10^{19} {\rm~cm^{-2}}$  in the direction of
LMC X--3. But, the allowed absorbing
column density beyond the LMC distance ($\lesssim 10^{19} {\rm~cm^{-2}}$) may 
still be consistent with the recent prediction
of a large-scale ($\sim 10^2$ kpc radius), very low density hot gaseous 
halo around the Milky Way, resulting from the accretion and cooling of the 
IGM  \citep{Mal04}. Additional contribution to \ovii\ may also
come from larger distances such as the Local Group, or even from the larger
scale local filament (Williams et al. 2005; Nicastro et al. 2002).
Furthermore, the hot gas properties 
might change from one sight-line to another; a comparison
between a couple of sight-lines may not be considered to be conclusive.
A systematic analysis of the existing data as well as sensitive new
observations would be very helpful. 

\section {Summary}

We have presented our TOO \chandra\ and \fuse\ spectroscopic observations 
of the black hole X-ray binary LMC X--3 in its high/soft state. 
These observations, enhanced by archival \fuse\ data,
allow us for the first time to measure both far-UV and 
X-ray absorption lines along a 50 kpc sight-line 
with little extragalactic confusion. Our main results and conclusions are
as follows:

\begin{itemize}

\item We have detected the absorption lines produced by the 
\ovii\ K$\alpha$ and \neix\ K$\alpha$ 
transitions. A joint fit of these lines, together with the upper bounds
to the \ovii\ K$\beta$, and \oviii\ K$\alpha$ lines, gives an 
\ovii\ column density of $\sim10^{16.3} {\rm~cm^{-2}}$
and a velocity dispersion of $\sim 80 {\rm~km~s^{-1}}$. These results are 
insensitive to the exact Ne/O elemental number ratio, to which only 
a 95\% confidence lower limit of 0.14 is determined. The hot gas, assumed 
to be in a CIE state, has a characteristic temperature in the range of $\sim
10^{6.1}$ K.

\item The X-ray-absorbing gas is likely to be primarily Galactic in origin.
This conclusion is based on the consistency of both the measured absolute 
\ovii\ line centroid
and its relative velocity offset from the \oi\ line, which should be 
predominantly Galactic. An origin of the gas in a wind 
from LMC X--3 may also be ruled out by the narrowness 
and apparent time invariance of the X-ray absorption 
lines. 

\item The \fuse\ spectrum shows a strong  ($\sim 6\sigma$ confidence) 
Galactic \ovi\ $\lambda 1031.93$  line absorption; 
there might be additional weak ($\sim 2\sigma$) 
\ovi\ absorption features at LMC velocities 
($\sim +310 {\rm~km~s^{-1}}$). The well-resolved Galactic absorption 
corresponds to an \ovi\ column density of $10^{14.5} 
{\rm~cm^{-2}}$, which is $\sim 2 \sigma$ greater than the value 
expected from the X-ray absorption line modeling.
This discrepancy may indicate a multi-phase state of the 
absorbing gas. 

\item We speculate that the highly-ionized absorbing 
gas represents a supernova-driven 
Galactic fountain, which globally resembles a thick gaseous disk with 
decreasing ionization away from the Galactic plane. In particular,
this Galactic fountain scenario naturally explains the 
significant nonthermal velocity broadening inferred from the direct \ovi\ line 
width measurement and from the relative strengths of the X-ray absorption 
lines. We further trace the warm \ion{H}{2} gas along the LMC X--3 sight-line
with our detected \oii\ K$\alpha$ line. From the X-ray absorption 
measurements and the Galactic (pulsar) dispersion measure in the LMC X--3 
vicinity, we estimate the oxygen abundance of the X-ray 
absorbing gas to be
$\gtrsim 16\% $ of the Galactic ISM value. Therefore, the X-ray-absorbing
gas is metally rich.

\item The X-ray absorption lines along the LMC X--3 sight-line 
have properties similar to those of other $cz \sim 0$ highly-ionized absorbers 
observed in the spectra of Galactic X-ray binaries and extragalactic 
AGNs (Yao \& Wang 2005). 
While there is still room for a significant presence of very 
low-density diffuse hot gas on scales  greater than 50 kpc (e.g., the Local 
Group), any such extragalactic studies of highly-ionized species at $z \sim 0$
must take into account the Galactic hot ISM component.
\end{itemize}

\acknowledgements We thank B. Savage for useful comments and the referee for 
a thorough report. The project was 
funded by NASA/SAO under the grant
G04-5046 and was based in part on an observation made with
NASA-CNES-CSA {\sl FUSE}, which 
is operated by the Johns Hopkins University under NASA contract
NAS5-32985. Funding for analysis of the {\it FUSE} data was provided
by NASA via grant NNG 04GJB83G. TMT appreciates additional support
from NASA LTSA grant NNG 04GG73G.

{}
\end{document}